\documentstyle[11pt,amssymb,epsfig,psfig]{article}

\begin{document}

\title{{\bf Casimir densities for a spherical shell in the global
monopole background }}

\author{
A. A. Saharian  $^{1}$\footnote{E-mail: saharyan@www.physdep.r.am
}
and  M. R. Setare $^2$ \footnote{E-mail: rezakord@yahoo.com} \\
 {$^1$ Department of Physics, Yerevan
State University, Yerevan, Armenia } \\
and
\\ {$^2$ Institute for Theoretical Physics and Mathematics, Tehran,
Iran} \\ {Department of Science, Physics Group, Kurdistan
University, Sanandeg, Iran}\\ {Department of Physics, Sharif
University of Technology, Tehran, Iran}}

\maketitle

\begin{abstract}
We investigate the vacuum expectation values for the
energy-momentum tensor of a massive scalar field with general
curvature coupling and obeying the Robin boundary condition on a
spherical shell in the $D+1$-dimensional global monopole
background. The expressions are derived for the Wightman function,
the vacuum expectation values of the field square, the vacuum
energy density, radial and azimuthal stress components in both
regions inside and outside the shell. A regularization procedure
is carried out by making use of the generalized Abel-Plana formula
for the series over zeros of cylinder functions. This formula
allows us to extract from the vacuum expectation values the parts
due to the global monopole gravitational field in the situation
without a boundary, and to present the boundary induced parts in
terms of exponentially convergent integrals, useful, in
particular, for numerical calculations. The asymptotic behavior of
the vacuum densities is investigated near the sphere surface and
at large distances. We show that for small values of the parameter
describing the solid angle deficit in global monopole geometry the
boundary induced vacuum stresses are strongly anisotropic.
\end{abstract}

\bigskip

{PACS number(s): 03.70.+k, 11.10.Kk}

\newpage

\section{Introduction}

The Casimir effect is one of the most interesting manifestations
of the nontrivial properties of the vacuum state in quantum field
theory. Since its first prediction by Casimir in 1948
\cite{Casi48} this effect has been investigated for various cases
of boundary geometries and different fields (see
\cite{Most97}--\cite{Milt02} and references therein).
Historically, the investigation of the Casimir effect for a
perfectly conducting spherical shell was motivated by the Casimir
semiclassical model of an electron. It has been shown by Boyer
\cite{Boye68} that the Casimir energy for the sphere is positive,
implying a repulsive force. This result was later reconsidered by
a number of authors \cite{Davi72}--\cite{Cand82}. More recently
new methods have been developed for this problem including direct
mode summation techniques based on the zeta function
regularization scheme \cite{Blau88}--\cite{Kirs01}. In Refs.
\cite{Seta01,Seta01b} the Casimir effect is considered for a
massless scalar field with the Dirichlet boundary condition on a
spherical shell with false/true vacuum inside/outside the shell.
Recently, the Casimir effect for spherical boundaries on de Sitter
and anti de Sitter bulks is also investigated
\cite{Noji00}--\cite{Eliz02}. In the investigations of the Casimir
effect the calculation of the local densities of the vacuum
characteristics is of special interest. In particular, these
include the vacuum expectation values of the energy-momentum
tensor.  In addition to describing the physical structure of the
quantum field at a given point, the energy-momentum tensor acts as
the source of gravity in the Einstein equations. It therefore
plays an important role in modeling a self-consistent dynamics
involving the gravitational field \cite{Birr82}. Investigation of
the energy distribution inside a perfectly reflecting spherical
shell was made in \cite{Olau81} in the case of QED and in
\cite{Olau81a} for QCD. The distributions of the other components
for the energy-momentum tensor of the electromagnetic field inside
as well as outside the shell, and in the region between two
concentric spherical shells are investigated in Refs.
\cite{Brev83}--\cite{Grig87}. Recently, in Ref. \cite{Saha01} the
vacuum expectation values of the energy-momentum tensor are
evaluated for a massive scalar field with general curvature
coupling parameter, satisfying the Robin boundary condition on
spherically symmetric boundaries in $D$-dimensional space.

The Casimir effect can be viewed as a polarization of vacuum by
boundary conditions. Another type of vacuum polarization arises in
the case of external gravitational field. In this paper we study a
situation when both types of sources for the polarization are
present. Namely, we consider a global monopole spacetime with a
concentric spherical shell. Topological defects have attracted a
great deal of attention because of their relevance to a number of
different areas ranging from condensed matter to structure
formation (for a review see \cite{Vile94}). In the context of hot
big bang cosmology, the unified theories of the fundamental
interactions predict that the universe passes through a sequence
of phase transitions. These phase transitions might have given
rise to several kinds of topological defects depending on the
nature of the symmetry that is broken \cite{Kibb76}. If a global
SO(3) symmetry of a triplet scalar field is broken, the point like
defects called global monopoles are believed to be formed. The
simplified global monopole was introduced by Sokolov and
Starobinsky \cite{Soko77}. The gravitational effects of the global
monopole were studied in Ref. \cite{Barr89}, where a solution is
presented which describes a global monopole at large radial
distances. The quantum vacuum effects of the matter fields on the
global monopole background have been considered in Refs.
\cite{Hisc90}--\cite{Beze00}. The effects produced by the non-zero
temperature are investigated as well \cite{Beze01temp}. The zeta
function and the Casimir energy for a spherical boundary in this
background are calculated in \cite{Bord96m,Beze00}.

In this paper the positive frequency Wightman function, the vacuum
expectation values of the field square and energy-momentum tensor
are investigated for a massive scalar field with general curvature
coupling parameter $\xi $, satisfying the Robin boundary condition
on a spherical shell in the $D+1$-dimensional spacetime of a
point-like global monopole. As special cases they include the
results for the Diriclet, Neumann, TM, and conformally invariant
Hawking boundary conditions. To evaluate the corresponding
bilinear field products we use the mode sum method in combination
with the summation formulae from Ref. \cite{Saha87} (see also
\cite{Saha00rev}). These formulae allow (i) to extract from vacuum
expectation values the parts due to the global monopole background
without boundaries, and (ii) to present the boundary induced parts
in terms of exponentially convergent integrals involving the
modified Bessel functions. We have organized the paper as follows.
In the next section we consider the vacuum inside a sphere and
derive formulae for the Wightman function, the expectation values
of the field square, energy density and stresses on the global
monopole background. The behavior of these quantities is
investigated near the boundary, at the sphere centre and for small
values of the parameter describing the solid angle deficit in the
global monopole geometry. Section \ref{msec2:out} is devoted to
the vacuum expectation values for the region outside a sphere.
Section \ref{msec3:conc} concludes the main results of the paper.

\section{Wightman function, vacuum expectation values of the
field square and energy-momentum tensor inside a spherical shell}
\label{secm:1}

\subsection{Wightman function}

Consider a real scalar field $\varphi $ with curvature coupling parameter $%
\xi $ on a $D+1$-dimensional global monopole background. In the
hyperspherical polar coordinates $(r,\vartheta ,\phi )\equiv
(r,\theta _{1},\theta _{2},\ldots \theta _{n},\phi )$, $n=D-2$,
the corresponding line element has the form
\begin{equation}
ds^{2}=dt^{2}-dr^{2}-\sigma ^{2}r^{2}d\Omega ^{2}_D,
\label{mmetric}
\end{equation}
where $d\Omega ^{2}_D$ is the line element on the surface of the
unit sphere in $D$-dimensional Euclidean space, the parameter
$\sigma $ is smaller than unity and is related to the symmetry
breaking energy scale in the theory. The solid angle corresponding
to Eq. (\ref{mmetric}) is $\sigma ^2S_D$ with $S_D=2\pi
^{D/2}/\Gamma (D/2)$ being the total area of the surface of the
unit sphere in $D$-dimensional Euclidean space. This leads to the
solid angle deficit $(1-\sigma ^2)S_D$ in the spacetime given by
line element (\ref{mmetric}). The field equation has the form
\begin{equation}
\left( \nabla _{i}\nabla ^{i}+m^{2}+\xi R\right) \varphi =0,\quad
\label{mfieldeq}
\end{equation}
where $R$ is the scalar curvature for the background space-time,
$m$ is the mass for the field quanta, $\nabla _{i}$ is the
covariant derivative operator associated with the metric
corresponding to line element (\ref{mmetric}). The values of the
curvature coupling parameter $\xi =0$, and $\xi =\xi _{D}$ with
$\xi _{D}\equiv (D-1)/4D$ correspond to the minimal and conformal
couplings, respectively. It is convenient for later use to write
the corresponding metric energy-momentum tensor in the form
\begin{equation}
T_{ik}=\nabla _{i}\varphi \nabla _{k}\varphi +\left[ \left( \xi -\frac{1}{4}%
\right) g_{ik}\nabla _{l}\nabla ^{l}-\xi \nabla _{i}\nabla _{k}-\xi R_{ik}%
\right] \varphi ^{2}.  \label{mTik}
\end{equation}
The nonzero components of the Ricci tensor and Ricci scalar for
the metric corresponding to line element (\ref{mmetric}) are given
by expressions
\begin{equation}
R_{2}^{2}=R_{3}^{3}=\cdots =R_{D}^{D}=n\frac{\sigma ^{2}-1}{\sigma
^2 r^{2}},\quad R=n(n+1)\frac{\sigma ^{2}-1}{\sigma ^2r^{2}},
\label{mRictens}
\end{equation}
where the indices $2,3,\ldots ,D$ correspond to the coordinates
$\theta _{1},\theta _{2},\ldots ,\phi $ respectively, and we adopt
the convention of Birrell and Davies \cite{Birr82} for the
curvature tensor. Note that for $\sigma \neq 1$ the geometry is
singular at the origin (point-like monopole), $r=0$.

In this paper we are interested in the vacuum expectation values
(VEVs) of the energy-momentum tensor (\ref{mTik}) on background of
the geometry described by (\ref{mmetric}), assuming that the field
satisfies the Robin boundary condition
\begin{equation}
\left( A_{1}+B_{1}n^{i}\nabla _{i}\right) \varphi (x)=0
\label{mrobcond}
\end{equation}
on the sphere of radius $a$, concentric with monopole. Here $A_{1}$ and $%
B_{1}$ are constants, and $n^{i}=(0,n^1,0,0)$ is the unit normal
to the sphere, $n^1=-1,1$ for the interior and exterior regions
respectively. The imposition of this boundary condition on the
quantum field $\varphi (x)$ leads to the modification of the
spectrum for the zero-point fluctuations and results in the shift
in VEVs for physical quantities. In particular, vacuum forces
arise acting on constraining boundary. This is the familiar
Casimir effect. By virtue of Eq. (\ref{mTik}) for the VEV of the
energy-momentum tensor we have
\begin{equation}
\langle 0|T_{ik}(x)|0\rangle =\lim_{x^{\prime }\rightarrow
x}\partial _{i}\partial _{k}^{\prime }\langle 0|\varphi (x)\varphi
(x^{\prime })|0\rangle +\left[ \left( \xi -\frac{1}{4}\right)
g_{ik}\nabla _{l}\nabla ^{l}-\xi \nabla _{i}\nabla _{k}-\xi
R_{ik}\right] \langle 0|\varphi ^{2}(x)|0\rangle ,
\label{mvevEMT}
\end{equation}
where $|0\rangle $ is the amplitude for the corresponding vacuum
state. Note that the expectation value $ \langle 0|\varphi
(x)\varphi (x^{\prime })|0\rangle \equiv G^{+}(x,x^{\prime })$ is
known as a positive frequency Wightman function. To derive the
expression for the regularized VEV of the field bilinear product
we will use the mode summation method. By expanding the field
operator over eigenfunctions and using the commutation rules one
can see that
\begin{equation}
\langle 0|\varphi (x)\varphi (x^{\prime })|0\rangle =\sum_{\alpha
}\varphi _{\alpha }(x)\varphi _{\alpha }^{\ast }(x^{\prime }),
\label{mfieldmodesum}
\end{equation}
where $\left\{ \varphi _{\alpha }(x),\varphi _{\alpha }^{\ast
}(x^{\prime })\right\} $ is a complete orthonormal set of positive
and negative frequency solutions to the field equation with
quantum numbers $\alpha $, satisfying boundary condition
(\ref{mrobcond}).

In the hyperspherical coordinates, for the region inside the
sphere the complete set of solutions to Eq. (\ref{mfieldeq}) with
scalar curvature from (\ref{mRictens}), has the form
\begin{equation}
\varphi _{\alpha }(x)=\beta _{\alpha }r^{-n/2}J_{\nu _{l}
}(\lambda r)Y(m_{k};\vartheta ,\phi )e^{-i\omega
t},\,\,l=0,1,2,\ldots , \label{meigfunc}
\end{equation}
where $m_{k}=(m_{0}\equiv l,m_{1},\ldots ,m_{n})$, and
$m_{1},m_{2},\ldots ,m_{n}$ are integers such that
\begin{equation}
0\leq m_{n-1}\leq m_{n-2}\leq \cdots \leq m_{1}\leq l,\quad
-m_{n-1}\leq m_{n}\leq m_{n-1},  \label{mnumbvalues}
\end{equation}
$J_{\nu }(z)$ is the Bessel function, and $Y(m_{k};\vartheta ,\phi
)$ is the
surface harmonic of degree $l$ (see \cite{Erdelyi}, Section 11.2). In Eq. (%
\ref{meigfunc}) we use the following notation
\begin{eqnarray}
\nu _{l}  &=&\frac{1}{\sigma }\left[ \left( l+\frac{n}{2}\right)
^{2}+(1-\sigma ^{2})n(n+1)\left( \xi -\xi _{D-1}\right) \right]
^{1/2},
\label{mnulam} \\
\lambda  &=&\sqrt{\omega ^{2}-m^{2}}.  \nonumber
\end{eqnarray}
In the following consideration we will assume that $\nu _{l} ^2$
is non-negative. This corresponds to the restriction on the values
of the curvature coupling parameter for $n>0$, given by the
condition
\begin{equation}\label{condposnu}
  \xi \geq -\frac{n}{4(n+1)(\sigma ^{-2}-1)}.
\end{equation}
Note that this condition is satisfied by the most important
special cases of the minimal and conformal couplings. The
coefficients $\beta _{\alpha }$ in Eq. (\ref{meigfunc}) can be
found from the normalization condition
\begin{equation}
\int \left| \varphi _{\alpha }(x)\right|
^{2}\sqrt{-g}dV=\frac{1}{2\omega }, \label{normcond}
\end{equation}
where the integration goes over the region inside the sphere.
Substituting eigenfunctions (\ref{meigfunc}), and using the
relation
\begin{equation}
\int \left| Y(m_{k};\vartheta ,\phi )\right| ^{2}d\Omega =N(m_{k})
\label{harmint}
\end{equation}
(the explicit form for $N(m_{k})$ is given in \cite{Erdelyi},
Section 11.3, and will not be necessary for the following
considerations in this paper) for the spherical harmonics and the
value for the standard integral involving the square of the Bessel
function, one finds
\begin{equation}
\beta _{\alpha }^{2}=\frac{\lambda T_{\nu _{l} }(\lambda a)}{%
N(m_{k})\omega a\sigma ^{D-1}},  \label{normcoef}
\end{equation}
with the notation
\begin{equation}
T_{\nu }(z)=\frac{z}{(z^{2}-\nu ^{2})J_{\nu }^{2}(z)+z^{2}J_{\nu
}^{^{\prime }2}(z)}.  \label{teka}
\end{equation}

From boundary condition (\ref{mrobcond}) on the sphere surface for
eigenfunctions (\ref{meigfunc}) one sees that the possible values
for the frequency have to be solutions to the following equation
\begin{equation}
AJ_{\nu _{l} }(z)+BzJ_{\nu _{l} }^{\prime }(z)=0,\quad z=\lambda
a,\quad A=A_{1}- \frac{n}{2}B,\quad B=\frac{n^1}{a}B_{1},
\label{eigenmodes}
\end{equation}
where $n^1=-1$ for the region inside the sphere. It is well known
(see, e.g., \cite{Erdelyi,Wats95}) that for real $A$, $B$, and
$\nu _{l}>-1$ all roots of this equation are simple and real,
except the case $A/B<-\nu _{l}$ when there are two purely
imaginary zeros. In the following we will assume values of $A/B$
for which all roots are real, $A/B\geq -\nu _{l}$. In terms of the
coefficients in (\ref{mrobcond}), it is sufficient to require
\begin{equation} \label{nu0}
  \frac{A_1a}{B_1}\leq \nu _0-\frac{n}{2}, \quad
  \nu _0=\frac{n}{2}\left[ 1+4(\sigma ^{-2}-1)
  \frac{n+1}{n}\xi  \right] ^{1/2 }.
\end{equation}
Note that this condition is satisfied for the most important
special cases of Dirichlet scalar, Neumann scalar ($A/B=1-D/2$)
with the curvature coupling parameter $\xi \geq 0$, and for a
scalar field with the TM type boundary condition ($A/B=D/2-1$).

Let $\lambda _{\nu _{l} ,k},\,k=1,2,\ldots ,$ be the positive
zeros of the function $AJ_{\nu _{l} }(z)+BzJ_{\nu _{l} }^{\prime
}(z)$, arranged in ascending order. The corresponding
eigenfrequencies $\omega =\omega _{\nu _{l} ,k}$ are related to
these zeros as $\omega _{\nu _{l} ,k}=\sqrt{\lambda _{\nu _{l}
,k}^{2}/a^{2}+m^{2}}$. Substituting Eq. (\ref{meigfunc}) into Eq.
(\ref{mfieldmodesum}) and using the addition formula  for the
spherical harmonics (see \cite{Erdelyi}, Section 11.4), one
obtains
\begin{eqnarray}
\langle 0|\varphi (x)\varphi (x^{\prime })|0\rangle
&=&\frac{(rr^{\prime })^{-n/2}}{naS_{D}\sigma
^{D-1}}\sum_{l=0}^{\infty }(2l+n)C_{l}^{n/2}(\cos \theta )
\label{fieldmodesum1} \\
&&\times \sum_{k=1}^{\infty }\frac{\lambda _{\nu _{l} ,k}T_{\nu
_{l} }(\lambda _{\nu _{l} ,k})}{\sqrt{\lambda _{\nu _{l}
,k}^{2}+m^{2}a^{2}}}J_{\nu _{l} }(\lambda _{\nu _{l} ,k}r/a)J_{\nu
_{l} }(\lambda _{\nu _{l} ,k}r^{\prime }/a)e^{i\omega _{\nu _{l}
,k}(t^{\prime }-t)}, \nonumber
\end{eqnarray}
where $C_{p}^{q}(x)$ is the Gegenbauer or ultraspherical
polynomial  of degree $p$ and order $q$, and $\theta $ is the
angle between directions $(\vartheta ,\phi )$ and $(\vartheta
^{\prime },\phi ^{\prime })$. To sum over $k$ we will use the
generalized Abel-Plana summation formula \cite{Saha87,Saha00rev}
\begin{eqnarray}
2\sum_{k=1}^{\infty }T_{\nu }(\lambda _{\nu ,k})f(\lambda _{\nu
,k}) &=&\int_{0}^{\infty }f(x)dx+\frac{\pi }{2}{\rm
Res}_{z=0}f(z)\frac{\bar{Y}
_{\nu }(z)}{\bar{J}_{\nu }(z)}{}  \nonumber \\
&-&\frac{1}{\pi }\int_{0}^{\infty }\ dx\frac{\bar{K}_{\nu
}(x)}{\bar{I}_{\nu }(x)}\left[ e^{-\nu \pi i}f(xe^{\pi
i/2})+e^{\nu \pi i}f(xe^{-\pi i/2}) \right] , \label{sumJ1anal}
\end{eqnarray}
where $I_{\nu }(x)$ and $K_{\nu }(x)$ are the modified Bessel
functions, and for a given function $F(z)$ we use the notation
\begin{equation}
\bar{F}(z)\equiv AF(z)+BzF^{\prime }(z).  \label{barnot}
\end{equation}
By taking in this formula $\nu =1/2$, $A=1$, $B=0$, as a
particular case we receive the Abel-Plana formula \cite{Hard49}.
Formula (\ref{sumJ1anal}) was applied previously to a number of
Casimir problems for spherically \cite{Grig86,Grig87,Saha01} and
cylindrically \cite{Saha88,Rome01} symmetric boundaries on the
Minkowski background, and for a wedge with and without circular
outer boundary \cite{Reza02}. Note that formula (\ref{sumJ1anal})
can be generalized in the case of the existence of purely
imaginary zeros for the function $\bar J_{\nu }(z)$ by adding the
corresponding residue term and taking the principal value of the
integral on the right (see Ref. \cite{Saha00rev}). As it has been
mentioned earlier, in this paper we will assume values of $A/B$
for which all solutions to Eq. (\ref{eigenmodes}) are real.
Applying to the sum over $k$ in Eq. (\ref{fieldmodesum1}) formula
(\ref{sumJ1anal}), the Wightman function can be presented in the
form
\begin{equation}
\langle 0|\varphi (x)\varphi (x^{\prime })|0\rangle =\langle 0_{{\rm m}%
}|\varphi (x)\varphi (x^{\prime })|0_{{\rm m}}\rangle +\langle
\varphi (x)\varphi (x^{\prime })\rangle _{b},  \label{unregWight1}
\end{equation}
where
\begin{equation}
\langle 0_{{\rm m}}|\varphi (x)\varphi (x^{\prime })|0_{{\rm
m}}\rangle = \frac{1}{2nS_{D}\sigma ^{D-1}}\sum_{l=0}^{\infty
}\frac{2l+n}{(rr^{\prime })^{n/2}} C_{l}^{n/2}(\cos \theta
)\int_{0}^{\infty }dz\,\frac{ze^{i\sqrt{z^{2}+m^{2}}(t^{\prime
}-t)}}{\sqrt{z^{2}+m^{2}}}J_{\nu _{l} }(zr)J_{\nu _{l}
}(zr^{\prime }), \label{Mink}
\end{equation}
and
\begin{eqnarray}
\langle \varphi (x)\varphi (x^{\prime })\rangle _{b}
&=&-\frac{1}{\pi naS_{D}\sigma ^{D-1} }\sum_{l=0}^{\infty
}\frac{2l+n}{(rr^{\prime })^{n/2}}C_{l}^{n/2}(\cos
\theta )\times   \label{Wightbound} \\
&&\int_{ma}^{\infty }dz\,z\frac{\bar{K}_{\nu _{l}
}(z)}{\bar{I}_{\nu _{l} }(z)} \frac{I_{\nu _{l} }(zr/a)I_{\nu _{l}
}(zr^{\prime }/a)}{\sqrt{z^{2}-m^{2}a^{2}}}\cosh \left[
\sqrt{z^{2}/a^{2}-m^{2}}(t^{\prime }-t)\right] . \nonumber
\end{eqnarray}
To obtain Eq. (\ref{Wightbound}) we have used the result that the
difference of the radicals is nonzero above the branch point only.
The conditions for the formula (\ref{sumJ1anal}) to be valid in
the case of the sum over $k$ in Eq. (\ref{fieldmodesum1}) are
satisfied if $r+r^{\prime }+|t-t^{\prime }|<2a$. The contribution
of the term (\ref{Mink}) to the VEV does not depend on $a$,
whereas the contribution of the term (\ref{Wightbound}) tends to
zero as $ a\rightarrow \infty $. It follows from here that
expression (\ref{Mink}) is the Wightman function for the unbounded
global monopole space with $|0_{{\rm m}}\rangle $ being the
amplitude for the corresponding vacuum. This can be seen also by
explicit evaluation of the mode sum using the eigenfunctions for
the global monopole spacetime without a boundary. Note that for
$\sigma =1$, $\nu _{l} $ is equal to $l+1/2$ and the sum over $l$
in Eq. (\ref{Mink}) can be summarized using the Gegenbauer
addition theorem for the Bessel function \cite{abramowiz}.
Evaluating the remaining integral involving the Bessel function
\cite{Prudnikov}, we obtain the standard expression for the
$D+1$-dimensional Minkowskian Wightman function. As we have seen,
the application of the generalized Abel-Plana formula allows us to
extract from the bilinear field product the contribution due to
the unbounded monopole spacetime, and the term (\ref{Wightbound})
can be interpreted as the part of the VEV induced by a spherical
boundary. In the massless case by using the formula for the
integral involving the Bessel functions, for the Wightman function
(\ref{Mink}) can be presented as
\begin{equation}
\langle 0_{{\rm m}}|\varphi (x)\varphi (x^{\prime })|0_{{\rm
m}}\rangle = \frac{1}{2\pi nS_{D}}\sum_{l=0}^{\infty
}\frac{(2l+n)C_{l}^{n/2}(\cos \theta )}{(\sigma ^{2}rr^{\prime
})^{(n+1)/2} }Q_{\nu _{l} -1/2}\left( \frac{r^{2}+r^{\prime
2}-(t-t^{\prime })^{2}+i\epsilon }{2rr^{\prime }}\right) ,
\label{Mink1}
\end{equation}
where $Q_{\nu _{l} }$ is the Legendre function of second kind, $\epsilon >0$ for $%
t>t^{\prime }$, and $\epsilon <0$ for $t<t^{\prime }$. Note that
similar expression for the Euclidean scalar Green function is
derived in Ref. \cite{Mazz91} for $D=3$ and in Ref. \cite{Beze02}
for an arbitrary $D$. The corresponding VEVs of the
energy-momentum tensor are investigated in Refs.
\cite{Hisc90}--\cite{Beze01temp}.

\subsection{VEV for the field square}

The VEV of the field square is obtained computing the Wightman
function in the coincidence limit $x^{\prime }\rightarrow x$. In
this limit expression (\ref{unregWight1}) gives a divergent result
and some renormalization procedure is needed. As we will see below
for $0<r<a$ the divergences are contained in the first summand on
the right of Eq. (\ref{unregWight1}) only. Hence, the
renormalization procedure for the local characteristics of the
vacuum, such as field square and energy-momentum tensor, is the
same as for the global monopole geometry without a boundary. This
procedure is discussed in a number of papers (see
\cite{Hisc90}--\cite{Beze00}) and is a useful illustration for the
general renormalization prescription on curved backgrounds (see,
for instance, \cite{Birr82}). In accordance with this
prescription, for the renormalization we must subtract from
(\ref{Mink}) the corresponding De Witt -- Schwinger expansion
involving the terms up to order $D$. For a massless field the
renormalized value of the field square has the following general
form
\begin{equation}\label{renphi2form}
\langle 0_{{\rm m}}|\varphi ^2(x)|0_{{\rm m}}\rangle _{{\rm
ren}}=\frac{1}{r^{D-1}}\left[ A^{(D)}(\sigma , \xi
)+B^{(D)}(\sigma ,\xi )\ln (\mu r)\right] ,
\end{equation}
where the arbitrary mass scale $\mu $ corresponds to the ambiguity
in the renormalization procedure. Note that this ambiguity is
absent for a spacetime of odd dimension and, hence, $B^{(D)}=0$
for even $D$. Because the dependence of the order of the Legendre
function in Eq. (\ref{Mink1}) on the parameter $\sigma $ is not a
simple one, it is not possible to obtain closed expression for the
coefficients $A^{(D)}$ and $B^{(D)}$. For small values $1-\sigma
^2$ approximate expressions are derived in Ref. \cite{Beze02} for
$D=4$ and $D=5$. In this paper our main interest are the parts in
the VEVs induced by the presence of a spherical shell and below we
will concentrate on these quantities.

Using the relation
\begin{equation}
C_{l}^{n/2}(1)=\frac{\Gamma (l+n)}{\Gamma (n)l!} , \label{Cl1}
\end{equation}
for the corresponding boundary part in the VEV of the field square
we get
\begin{equation}
\langle \varphi ^{2}(x)\rangle _{b}=-\frac{1}{\pi
ar^{n}S_{D}\sigma ^{D-1}} \sum_{l=0}^{\infty }D_{l}\int_{ma
}^{\infty }dz\ \frac{\bar{K}_{\nu _{l} }(z)}{\bar{I}_{\nu _{l}
}(z)}\frac{zI_{\nu _{l} }^{2}(zr/a)}{\sqrt{z^{2}-m^{2}a^{2}}},
\label{regsquare}
\end{equation}
where
\begin{equation}
D_{l}=(2l+D-2)\frac{\Gamma (l+D-2)}{\Gamma (D-1)\,l!}
\label{Dlang}
\end{equation}
is the degeneracy of each angular mode with given $l$. From the
well known properties of the modified Bessel functions it follows
that in the case of a Dirichlet scalar this quantity is negative
and for $\xi \geq 0$ is a monotonic decreasing function on $r$. As
it has been noted earlier, expression (\ref{regsquare}) is finite
for all values $0<r<a$. For a given $l$ and large $z$ the
subintegrand behaves as $e^{2z(r/a-1)}/z$ and, hence, the integral
converges when $r<a$. For large values $l$, introducing a new
integration variable $y=z/\nu _{l} $ in the integral of Eq.
(\ref{regsquare}) and using the uniform asymptotic expansions for
the modified Bessel functions \cite{abramowiz}, it can be seen
that to the leading order over $l$ the subintegrand multiplied by
$D_l$ behaves as
\begin{equation}\label{asympsubint}
  l^n\frac{\exp\left\{ 2\nu _{l} \left[ \eta (zr/a)-\eta (z)\right]
  \right\}}{\sqrt{1+(zr/a)^2}}, \quad \eta (z)=\sqrt{1+z^2}+\ln
  \frac{z}{1+\sqrt{1+z^2}},
\end{equation}
and, hence, the both integral and sum are convergent for $r<a$ and
diverge at $r=a$. For the points near the sphere the leading term
of the corresponding asymptotic expansion over $1/(a-r)$ has the
form
\begin{equation}\label{asympleadterm}
\langle \varphi ^{2}(x)\rangle _{b}\approx \frac{(1-2\delta
_{B0})\Gamma \left( \frac{D-1}{2}\right)}{(4\pi
)^{(D+1)/2}(a-r)^{D-1}}.
\end{equation}
This term does not depend on the mass and parameter $\sigma $ and
is the same as that for a sphere on the Minkowski bulk. As the
purely gravitational part (\ref{renphi2form}) is finite for $r=a$,
we see that near the sphere surface the VEV of the field square is
dominated by the boundary induced part, Eq. (\ref{regsquare}).

In the limit $r\to 0$ the main contribution into Eq.
(\ref{regsquare}) comes from the $l=0$ summand with the leading
term
\begin{equation}\label{limitrto0}
\langle \varphi ^{2}(x)\rangle _{b}\approx -\frac{\Gamma ^{-2}(\nu
_0+1)}{2^{2\nu _0}\pi a ^{D-1}S_D\sigma ^{D-1}}\left(
\frac{r}{a}\right) ^{2\nu _0-n}\int _{ma}^{\infty }dz
\frac{z^{2\nu _0+1}}{\sqrt{z^2-m^2a^2}}\frac{\bar K_{\nu
_0}(z)}{\bar I_{\nu _0}(z)},\quad r\to 0,
\end{equation}
where $\nu _0$ is given by formula (\ref{nu0}). As a result the
boundary induced VEV (\ref{regsquare}) is zero at the sphere
centre for $(\sigma ^{-2}-1)\xi >0$, non-zero constant for
$(\sigma ^{-2}-1)\xi =0$, and is infinite for $(\sigma ^{-2}-1)\xi
<0$. As it follows from expressions (\ref{renphi2form}) and
(\ref{limitrto0}), for $r\ll a $ the total VEV is dominated by the
purely gravitational part.

\begin{figure}[tbph]
\begin{center}
\begin{tabular}{ccc}
\epsfig{figure=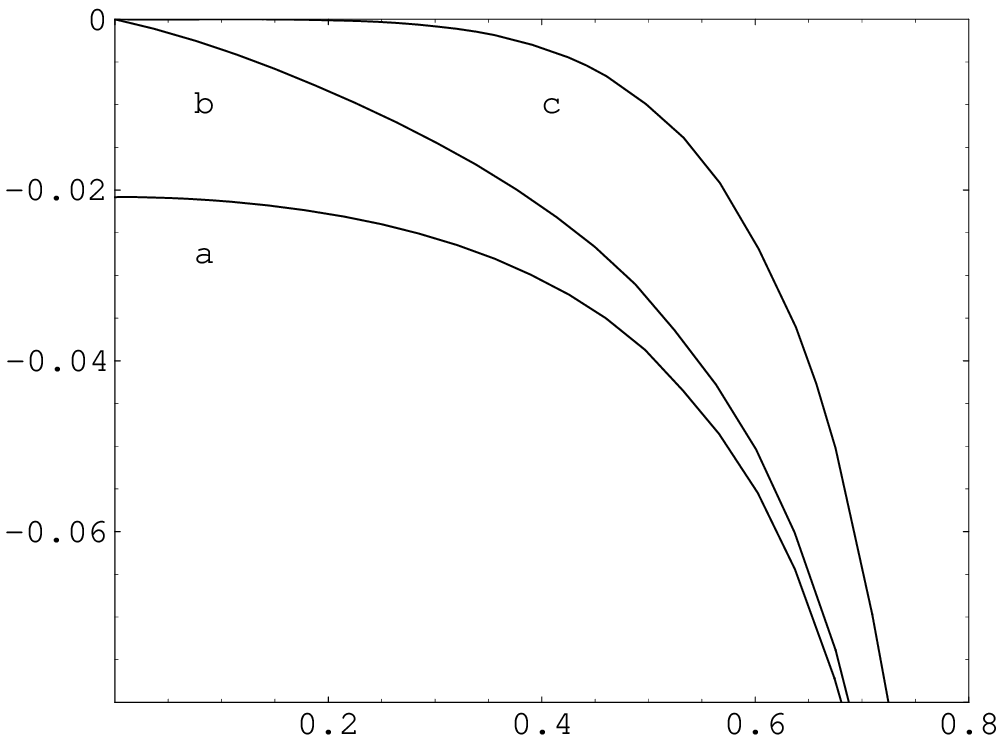,width=6cm,height=5cm} & \hspace*{0.5cm} & %
\epsfig{figure=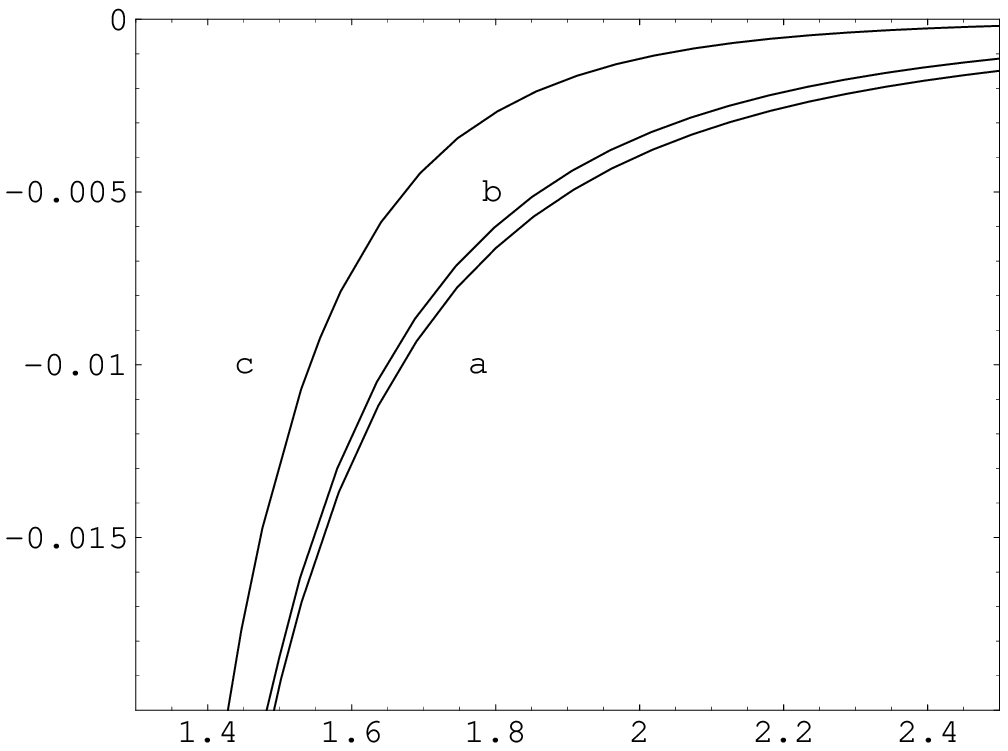,width=6cm,height=5cm}
\end{tabular}
\end{center}
\caption{ The sphere induced VEV $a^{D-1}\langle \varphi
^{2}(x)\rangle _{b}$ as a function on $r/a$ in the case of a
conformally coupled massless $D=3$ Dirichlet scalar for the
regions inside (left panel) and outside (right panel) the sphere.
The curves are plotted for $\sigma =1$ (a), $\sigma =0.5$ (b),
$\sigma =0.2$ (c)} \label{fig1phi2}
\end{figure}

Now we turn to the limit $\sigma \ll 1$ for a fixed value $r<a$.
To satisfy condition (\ref{condposnu}) we will assume that $\xi
\geq 0$. For $\xi >0$, from Eq. (\ref{mnulam}) one has $\nu _{l}
\gg 1$, and after introducing in Eq. (\ref{regsquare}) a new
integration variable $y=z/\nu _{l} $, we can replace the modified
Bessel function by their uniform asymptotic expansions for large
values of the order. The integral over $y$ can be estimated by
using the Laplace method. The main contribution to the sum over
$l$ comes from the summand with $l=0$, and to the leading order we
receive
\begin{equation}\label{phi2sigll1}
\langle \varphi ^{2}(x)\rangle _{b}\approx \frac{(1-2\delta
_{B0})\exp \left[-2\tilde \nu \ln (a/r)\right]}{(8\pi \tilde \nu
)^{1/2}r^{n}S_D\sigma ^{D-1}\sqrt{a^2-r^2}},\quad \tilde \nu
=\frac{1}{\sigma }\sqrt{n(n+1)\xi },\quad \sigma \ll 1.
\end{equation}
For $\xi =0$ and $\sigma \ll 1$ for the terms with $l\neq 0$ one
has $\nu _{l} \gg 1$. The corresponding contribution can be
estimated by the way similar to that in the previous case. This
contribution is exponentially small. For the summand with $l=0$ to
the leading order over $\sigma $ we have $\nu _{l} =n/2$ and,
hence,
\begin{equation}\label{phi2sigll1b}
\langle \varphi ^{2}(x)\rangle _{b}\approx -\frac{1}{\pi a
r^nS_D\sigma ^{D-1}}\int _{ma}^{\infty }dz \frac{\bar
K_{n/2}(z)}{\bar
I_{n/2}(z)}\frac{zI^2_{n/2}(zr/a)}{\sqrt{z^2-m^2a^2}},\quad \xi
=0,\quad \sigma \ll 1.
\end{equation}
Note that for odd values $n$ the modified Bessel functions in this
formula are expressed via the elementary functions. In Fig.
\ref{fig1phi2} (left panel) we have plotted the dependence of the
boundary induced VEV (\ref{regsquare}) on $r/a$ for the region
inside the sphere in the case of a conformally coupled ($\xi =\xi
_D $) massless Dirichlet scalar in $D=3$ spatial dimensions. The
separate curves correspond to the values $\sigma =1$ (a), $\sigma
=0.5$ (b), $\sigma =0.2$ (c).

\subsection{Vacuum energy-momentum tensor inside a sphere}

Substituting the Wightman function (\ref{unregWight1}) with
expressions (\ref{Mink}) and (\ref{Wightbound}) into Eq.
(\ref{mvevEMT}), for the VEV of the energy-momentum tensor inside
a spherical shell one finds
\begin{equation}
\langle 0|T_{i}^{k}|0\rangle ={\rm diag}\left( \varepsilon
,-p,-p_{\perp },\ldots ,-p_{\perp }\right) ,  \label{diagEMT}
\end{equation}
where the vacuum energy density $\varepsilon $ and the effective
pressures in radial, $p$, and azimuthal, $p_{\perp }$, directions
are functions of the radial coordinate only. Similar to the
Wightman function, the components of the vacuum energy-momentum
tensor can be presented in the form
\begin{equation}\label{compdecomp}
  q=q_m+q_b,\quad q=\varepsilon ,\,p,\,p_{\perp },
\end{equation}
where $q_m$ are the corresponding quantities for the monopole
geometry when the boundary is absent, and quantities $q_b$ are
induced by the presence of the spherical shell. As a consequence
of the continuity equation $\nabla _{k}\langle
0|T_{i}^{k}|0\rangle =0$, these functions are related by the
equation
\begin{equation}
r\frac{dp_i}{dr}+(D-1)(p_i-p_{\perp i})=0,\quad i=m,b.
\label{conteq}
\end{equation}
For massless fields the VEV of the energy-momentum tensor for the
global monopole geometry without boundaries are investigated in
Refs. \cite{Hisc90}--\cite{Beze02}. The corresponding renormalized
components have the structure similar to that given in Eq.
(\ref{renphi2form}) for the field square:
\begin{equation}\label{renEMTpuremon}
q_m =\frac{1}{r^{D+1}}\left[ q_m^{(1)}+q_m^{(2)}\ln (\mu r)\right]
,
\end{equation}
where the coefficients $q_m^{(1)}$, $q_m^{(2)}$ depend only on the
parameters $\sigma $ and $\xi $, and $q_m^{(2)}=0$ in even
dimensions $D$. From equation (\ref{conteq}) the following
relations between these coefficients are obtained
\begin{equation}\label{relcoef}
p_m^{(1)}=-\frac{D-1}{2}p_{\perp
m}^{(1)}+\frac{1}{2}p_m^{(2)},\quad
p_m^{(2)}=-\frac{D-1}{2}p_{\perp m}^{(2)}.
\end{equation}
In particular, for conformally coupled fields all coefficients
$q_m^{(1)}$, $q_m^{(2)}$ can be expressed via $\varepsilon
_m^{(1)}$, $\varepsilon _m^{(2)}$ using the trace anomaly.

From Eqs. (\ref{mvevEMT}), (\ref{Wightbound}), and
(\ref{regsquare}) for the boundary induced parts of the
energy-momentum tensor components one obtains
\begin{equation}
q_b(a,r)=-\frac{1}{2\pi a^{3}r^{n}S_{D}\sigma
^{D-1}}\sum_{l=0}^{\infty
}D_{l}\int_{ma}^{\infty }dz\,z^{3}\frac{\bar{K}_{\nu _{l} }(z)}{\bar{I}_{\nu _{l} }(z)}%
\frac{F_{\nu _{l} }^{(q)}\left[ I_{\nu _{l} }(zr/a)\right] }{\sqrt{z^{2}-m^{2}a^{2}}}%
,\quad r<a,  \label{q1in}
\end{equation}
where for a given function $f(y)$ we have introduced the notations
\begin{eqnarray}
F_{\nu _{l} }^{(\varepsilon )}\left[ f(y)\right]  &=&(1-4\xi
)\left[ f^{^{\prime
}2}(y)-\frac{n}{y}f(y)f^{\prime }(y)+\left( \frac{\nu _{l} ^{2}}{y^{2}}-\frac{%
1+4\xi -2(mr/y)^{2}}{1-4\xi }\right) f^{2}(y)\right]   \label{Fineps} \\
F_{\nu _{l} }^{(p)}\left[ f(y)\right]  &=&f^{^{\prime }2}(y)+\frac{\tilde \xi }{y}%
f(y)f^{\prime }(y)-\left( 1+\frac{\nu _{l} ^{2}+\tilde \xi n/2}{y^{2}}%
\right) f^{2}(y),\; \tilde \xi =4(n+1)\xi -n  \label{Finperad} \\
F_{\nu _{l} }^{(p_{\perp })}\left[ f(y)\right]  &=&(4\xi -1)f^{^{\prime }2}(y)-%
\frac{\tilde \xi }{y}f(y)f^{\prime }(y)+\left[ 4\xi -1+\frac{\nu
_{l} ^{2}(1+\tilde \xi )+\tilde \xi n/2}{(n+1)y^{2}}\right]
f^{2}(y). \label{Finpeaz}
\end{eqnarray}
It can be seen that components (\ref{q1in}) satisfy Eq.
(\ref{conteq}) and are finite for $0<r<a$.

In the case $D=1$ the angular part in line element (\ref{mmetric})
is absent and we obtain VEVs for the one -- dimensional segment
$-a\leq x\leq a$. In this case $\langle
0_{{\mathrm{m}}}|T_i^k|0_{{\mathrm{m}}}\rangle _{{\mathrm{ren}}}=0
$ and the boundary part remains only. Due to the gamma function in
the denominator of the expression for $D_{l}$ in Eq.
(\ref{Dlang}), now the only nonzero coefficients are $
D_{0}=D_{1}=1$. Using the standard expressions for the functions
$I_{\pm 1/2}(z)$ and $K_{\pm 1/2}(z)$ we obtain the formulae for
the vacuum energy-momentum tensor given in Refs.
\cite{Saha01,Rome02}. Note that in this case the vacuum stresses
are uniform and do not depend on the parameter $\xi $. For $D=2$
and $r\neq 0$ the line element (\ref{mmetric}) corresponds to the
flat background spacetime and coincides with the $2D$ cosmic
string geometry. Note that in this case $n=0$ and from Eq.
(\ref{mnulam}) one has $\nu _{l} =l/\sigma $.

At the sphere center the main contribution to the boundary induced
VEV (\ref{q1in}) comes from the summands with $l=0$ and one has
\begin{equation}\label{compatcentre}
  q_b(a,r)\sim -\frac{(2\nu _0-n)r^{2\nu _0-n-2}f^{(q)}_{0}}{
  (2a)^{2\nu _0+1}\pi S_D\sigma ^{D-1}\Gamma ^2(\nu _0+1)}
  \int _{ma}^{\infty }dz \frac{z^{2\nu _0+1}}{\sqrt{z^2-m^2a^2}}
  \frac{\bar K_{\nu _0}(z)}{
  \bar I_{\nu _0}(z)},\quad r\to 0,
\end{equation}
where the following notations are introduced
\begin{equation}\label{fnu0}
  f^{(\varepsilon )}_{0}=(1-4\xi )\nu _0,\quad f^{(p)}_{0}=
 \frac{\tilde \xi }{2},\quad f^{(p_{\perp })}_{0}=\frac{\nu
 _0-1/2}{D-1}\tilde \xi \label{Fq0} .
\end{equation}
It follows from Eq. (\ref{compatcentre}) that the boundary induced
components are zero at the sphere centre for $n\xi (\sigma
^{-2}-1)> 1$, are non-zero constants for $n\xi (\sigma ^{-2}-1)=
1$, and are infinite otherwise. These singularities appear because
the geometrical characteristics of global monopole spacetime  are
divergent at the origin. However, note that the corresponding
contribution to the total energy of the vacuum inside a sphere
coming from $\varepsilon _b$ is finite due to the factor $r^{D-1}$
in the volume element. Comparing expressions (\ref{renEMTpuremon})
and (\ref{compatcentre}) we conclude that near the centre the
vacuum energy-momentum tensor is dominated by the purely
gravitational part.

Expectation values (\ref{q1in}) diverge at the sphere surface, $
r\rightarrow a$. These divergencies are well-known in quantum
field theory with boundaries and are investigated for various
types of boundary geometries. For the problem under consideration
the corresponding asymptotic behavior can be found using the
uniform asymptotic expansions for the modified Bessel functions,
and the leading terms are determined by the relations
\begin{equation}
p_{b\perp }\sim -\varepsilon _b\sim \frac{D a p_b}{(D-1)(a-r)}
\sim \frac{D\Gamma ((D+1)/2)(\xi -\xi _{D})}{2^{D}\pi
^{(D+1)/2}(a-r)^{D+1}}\left( 1-2\delta _{B0}\right) .
\label{epsinasimp}
\end{equation}
These terms do not depend on mass and parameter $\sigma $, and are
the same as for a spherical shell in the Minkowski bulk (see, for
instance, \cite{Saha01}). For a conformally coupled scalar the
coefficients for the leading terms are zero and $\varepsilon
,p_{\perp }\sim (a-r)^{-D}$, $p\sim (a-r)^{1-D}$. In general,
asymptotic series can be developed in powers of the distance from
the boundary. The corresponding subleading coefficients will
depend on the mass, Robin coefficient, and parameter $\sigma $.
Due to surface divergencies near the sphere surface the total
vacuum energy-momentum tensor is dominated by the boundary induced
parts $q_b$.

\begin{figure}[tbph]
\begin{center}
\begin{tabular}{ccc}
\epsfig{figure=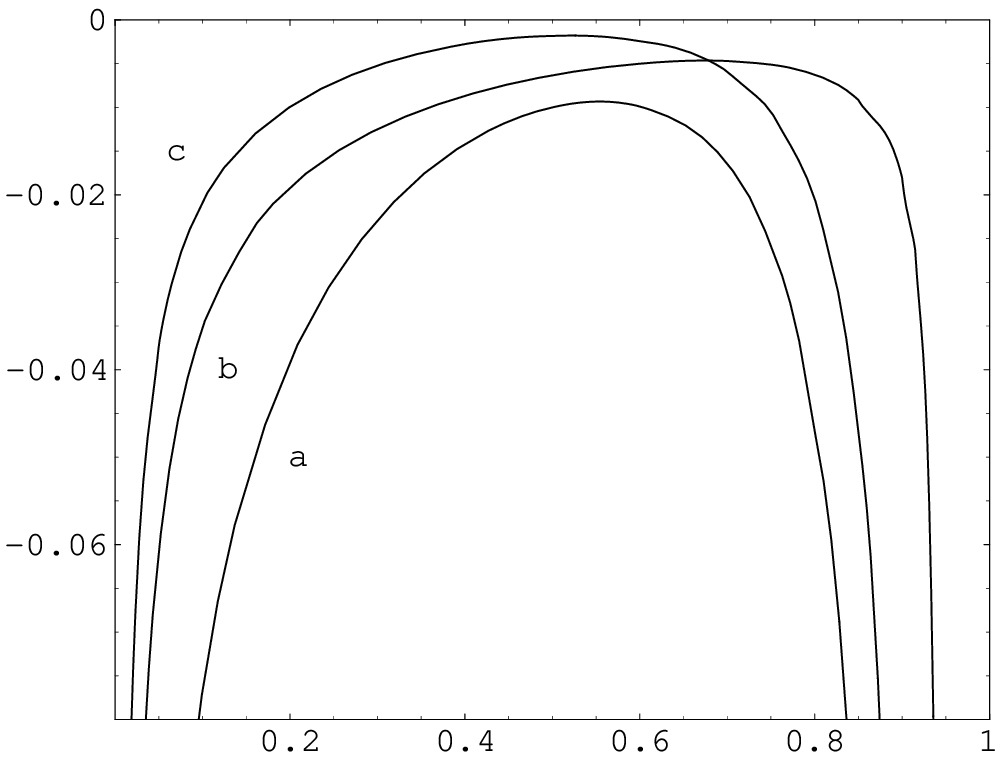,width=6cm,height=5cm} & \hspace*{0.5cm} & %
\epsfig{figure=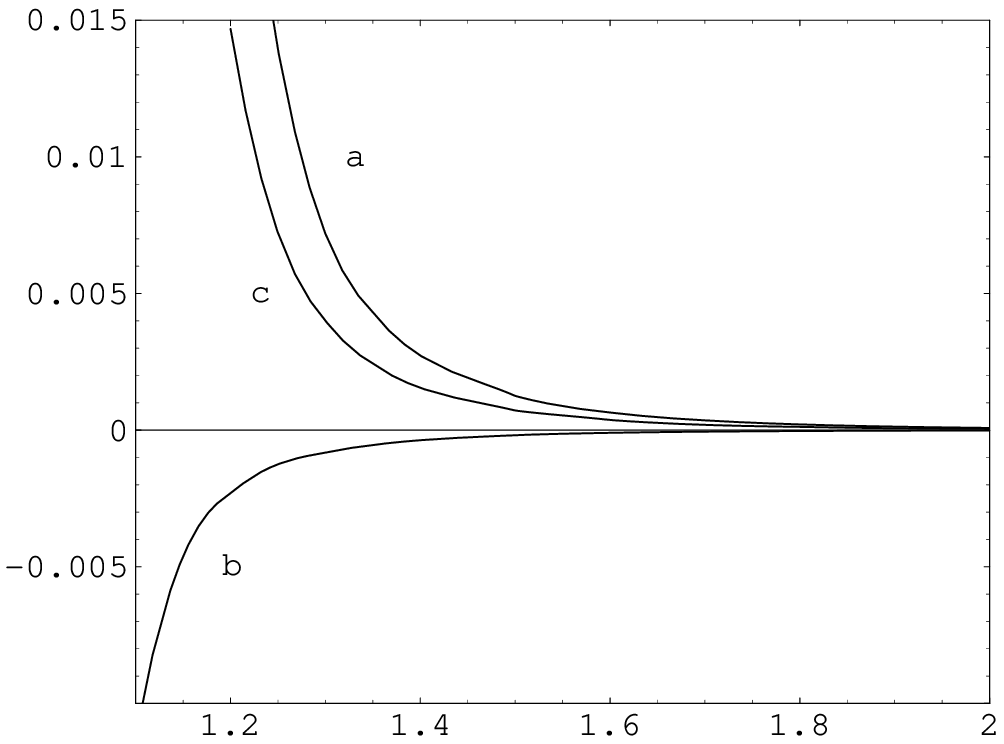,width=6cm,height=5cm}
\end{tabular}
\end{center}
\caption{ The sphere induced VEVs of the energy density,
$\varepsilon _b$ (curves a), radial pressure, $p_b$ (curves b) and
azimuthal pressure, $p_{\perp b}$ (curves c) multiplied by
$a^{D+1}$, as functions on $r/a$ in the case of a conformally
coupled massless $D=3$ Dirichlet scalar for the regions inside
(left panel) and outside (right panel) the sphere. The curves are
plotted for $\sigma =0.5$. } \label{fig2EMT}
\end{figure}

Now let us consider the VEVs of the energy-momentum tensor in the
limit $\sigma \ll 1$ for a fixed $r<a$. For $\xi >0$ by the
calculations similar to those given above for the field square,
for the corresponding boundary parts one receives
\begin{equation}\label{EMTsigll1}
  q_b(a,r)\approx \frac{(1-2\delta _{B0})\tilde \nu ^{3/2}\exp\left[ -2
  \tilde \nu \ln(a/r)\right] }{(8\pi )^{1/2}S_D\sigma
  ^{D-1}r^D\sqrt{a^2-r^2}}f_1^{(q)},
\end{equation}
where $\tilde \nu $ is defined in Eq. (\ref{phi2sigll1}), and
\begin{equation}\label{f1q}
f_1^{(\varepsilon )}=1-4\xi ,\quad f_1^{(p)}=\frac{\tilde
\xi}{2\tilde \nu },\quad f_1^{(p_{\perp })}=\frac{\tilde
\xi}{D-1}.
\end{equation}
Note that in this case the boundary induced vacuum stresses are
strongly anisotropic: $p_b/p_{\perp b}\sim \sigma \ll 1$. For a
minimally coupled scalar , $\xi =0$, the leading term of the
asymptotic expansion over $\sigma $ comes from the $l=0$ summand
in Eq. (\ref{q1in}) with $\nu _{l} =n/2$. This term behaves as
$\sigma ^{1-D}$. In Fig. \ref{fig2EMT} (left panel) we have
presented the quantities $\varepsilon _b$, $p_b$, $p_{\perp b}$
inside a spherical shell as functions on $r/a$ in the case of a
conformally coupled massless $D=3$ Dirichlet scalar on background
of the global monopole with $\sigma =0.5$. Note that for these
values of parameters $n\xi (\sigma ^{-2}-1)<1$, and the vacuum
energy-momentum tensor components are infinite at the sphere
centre.

\section{Vacuum densities outside a sphere}
\label{msec2:out}

To obtain the VEV for the energy-momentum tensor outside a sphere
we consider first the scalar
vacuum in the layer between two concentric spheres with radii $a$ and $b$, $%
a<b$. The boundary conditions on these surfaces we will take in
the form
\begin{equation}
\left( A_{1}+B_{1}\frac{\partial }{\partial r}\right) \varphi
(x)=0,\quad r=a,b.  \label{twospherebc}
\end{equation}
The corresponding eigenfunctions can be obtained from Eq.
(\ref{meigfunc}) with the replacement
\begin{equation}
J_{\nu _{l} }(\lambda r)\rightarrow g_{\nu _{l} }(\lambda
a,\lambda r)\equiv J_{\nu _{l} }(\lambda r)\bar{Y}_{\nu _{l}
}(\lambda a)-\bar{J}_{\nu _{l} }(\lambda a)Y_{\nu _{l} }(\lambda
r), \label{genu}
\end{equation}
where $Y_{\nu _{l} }(z)$ is the Neumann function, and the
functions with overbars are defined in accordance with Eq.
(\ref{barnot}). Note that now in the definition of the
coefficients in formula (\ref{eigenmodes}) we have to take
$n^1=1$. The functions chosen in this way satisfy the boundary
condition on the sphere $r=a$. From the boundary condition on
$r=b$ one obtains that the corresponding eigenmodes are solutions
to the equation
\begin{equation}
C_{\nu _{l} }^{ab}(\eta ,\lambda a)\equiv \bar{J}_{\nu _{l}
}(\lambda a)\bar{Y}^{(b)}_{\nu _{l} }(\lambda
b)-\bar{J}^{(b)}_{\nu _{l} }(\lambda b)\bar{Y}_{\nu _{l} }(\lambda
a)=0, \quad \eta =b/a, \label{eigmodesab}
\end{equation}
with the notation
\begin{equation}\label{Fb}
  \bar{F}^{(b)}(z)=A_bF(z)+B_bzF'(z),\quad
  A_b=A_1-\frac{B_1n}{2b}, \quad B_b=\frac{B_1}{b},
\end{equation}
for a given function $F(z)$. In Appendix \ref{sec:app1} we show
that for large values $\eta $ all roots to Eq. (\ref{eigmodesab})
are real. As we are interested in the VEVs for the region outside
a single sphere, obtained in the limit $b\to \infty $, this is the
case for our consideration below. Let $\gamma _{\nu _{l}
,k}=\lambda a$ , $ k=1,2,\ldots $ be the corresponding positive
roots, arranged in ascending order. The coefficients $\beta
_{\alpha }$ are determined from the normalization condition
(\ref{normcoef}), where now the integration goes over the region
between the spheres, $a\leq r\leq b$. Using the standard formula
for integrals involving the product of any two cylinder functions
one obtains
\begin{equation}
\beta _{\alpha }^{2}=\frac{\pi ^{2}\lambda }{4N(m_{k})\omega a\sigma ^{D-1}}%
T_{\nu _{l} }^{ab}(\eta ,\lambda a),  \label{betanormab}
\end{equation}
where $N(m_{k})$ comes from the normalization integral
(\ref{harmint}) and we use the notation
\begin{equation}
T_{\nu _{l} }^{ab}(\eta ,z)=z\left\{ \frac{\bar{J}_{\nu _{l}
}^{2}(z)}{\bar{J}_{\nu _{l} }^{2}(\eta z)}\left[
A^{2}_{b}+B^{2}_{b}(\eta ^{2}z^{2}-\nu _{l} ^{2})\right]
-A^{2}-B^{2}(z^{2}-\nu _{l} ^{2})\right\} ^{-1}, \label{tekaAB}
\end{equation}
with constants $A$, $B$, $A_b$, $B_b$ defined in
(\ref{eigenmodes}) and (\ref{Fb}).

Substituting the eigenfunctions into the mode sum
(\ref{mfieldmodesum}) and using the addition formula for the
spherical harmonics, the expectation value of the field product is
presented in the form
\begin{equation}
\langle 0|\varphi (x)\varphi (x^{\prime })|0\rangle =\frac{\pi
^{2}(rr^{\prime })^{-n/2}}{4naS_{D}\sigma
^{D-1}}\sum_{l=0}^{\infty }(2l+n)C_{l}^{n/2}(\cos \theta
)\sum_{k=1}^{\infty }h(\gamma _{\nu _{l} ,k})T_{\nu _{l}
}^{ab}(\eta ,\gamma _{\nu _{l} ,k}),  \label{fieldmodesum1ab}
\end{equation}
with the function
\begin{equation}
h(z)=\frac{ze^{i\sqrt{z^{2}/a^{2}+m^{2}}(t^{\prime
}-t)}}{\sqrt{z^{2}+m^{2}a^{2}}}g_{\nu _{l} }(z,zr/a)g_{\nu _{l}
}(z,zr^{\prime }/a). \label{hab}
\end{equation}
To sum over $k$ we will use the summation formula
\cite{Saha01,Saha87,Saha00rev}
\begin{eqnarray}
\frac{\pi ^{2}}{2}\sum_{k=1}^{\infty }h(\gamma _{\nu ,k})T_{\nu
}^{ab}(\eta ,\gamma _{\nu ,k})&=&\int_{0}^{\infty
}\frac{h(x)dx}{\bar{J}_{\nu
}^{2}(x)+\bar{Y}_{\nu }^{2}(x)} -r_{\nu }[h(z)] \label{cor3form} \\
&&-\frac{\pi }{4}\int_{0}^{\infty }\frac{\bar{K}_{\nu }^{(b)}(\eta x)}{\bar{K}%
_{\nu }(x)}\frac{\left[ h(xe^{\pi i/2})+h(xe^{-\pi i/2})\right] dx}{\bar{K}%
_{\nu }(x)\bar{I}_{\nu }^{(b)}(\eta x)-\bar{K}_{\nu }^{(b)}(\eta
x)\bar{I}_{\nu }(x)}, \nonumber
\end{eqnarray}
where we have assumed that all zeros for the function
(\ref{eigmodesab}) are real. In this formula the term
\begin{equation}\label{rnu}
r_{\nu }[h(z)]=\pi \sum_{k,s=1,2}{\mathrm{Res}}_{z=z_{\nu ,
k}^{(s)}} \left[ \frac{\bar{H}_{\nu }^{(sb)}(\eta
z)h(z)}{\bar{H}_{\nu }^{(s)}(z)C^{ab}_{\nu }(\eta ,z)}\right] =\pi
i\sum_{k,s=1,2} \frac{(-1)^{s+1}h(z_{\nu ,k}^{(s)})}{\bar{H}_{\nu
}^{(s)'}(z_{\nu ,k}^{(s)})\bar{J}_{\nu }(z_{\nu ,k}^{(s)})}
\end{equation}
comes from the residues at the possible zeros $z=z_{\nu ,k}^{(s)}$
of the function $\bar{H}^{(s)}_{\nu }(z)$, $s=1,2$, with
$0<{\mathrm{arg}} z_{\nu ,k}^{(1)}<\pi /2$, $z_{\nu ,
k}^{(2)}=z_{\nu ,k}^{(1)*}$. Here and below $H_{\nu }^{(s)}(z)$,
$s=1,2$ are the Hankel functions. Note that, as we have shown in
Appendix \ref{sec:app1}, under the condition (\ref{nu0}) the
function  $\bar{K}_{\nu }(z)$ has no real zeros and, hence, the
function $\bar{H}^{(s)}_{\nu }(z)$ has no purely imaginary zeros.
In the case of function (\ref{hab}) the corresponding conditions
for (\ref{cor3form}) are satisfied if $r+r^{\prime }+|t-t^{\prime
}|<2b$. Note that this is the case in the coincidence limit for
the region under consideration. Applying to the sum over $k$ in
Eq. (\ref{fieldmodesum1ab}) formula (\ref{cor3form}), for the
corresponding Wightman function one obtains
\begin{equation}
\langle 0|\varphi (x)\varphi (x^{\prime })|0\rangle
=\frac{1}{2naS_{D}\sigma ^{D-1}}\sum_{l=0}^{\infty
}\frac{2l+n}{(rr^{\prime })^{n/2}}C_{l}^{n/2}(\cos
\theta )\left\{ \int_{0}^{\infty }\frac{h(z)dz}{\bar{J}_{\nu _{l} }^{2}(z)+\bar{Y}%
_{\nu _{l} }^{2}(z)}-r_{\nu _l}[h(z)]\right.
\label{unregWightab}
\end{equation}
\[
-\left. \frac{2}{\pi }\int_{ma}^{\infty }\frac{zdz}{\sqrt{z^{2}-a^{2}m^{2}}}%
\frac{\bar{K}_{\nu _{l} }^{(b)}(\eta z)}{\bar{K}_{\nu _{l}
}(z)}\frac{G_{\nu _{l} }(z,zr/a)G_{\nu _{l} }(z,zr^{\prime
}/a)}{\bar{K}_{\nu _{l} }(z)\bar{I}_{\nu _{l} }^{(b)}(\eta
z)-\bar{K}_{\nu _{l} }^{(b)}(\eta z)\bar{I}_{\nu _{l} }(z)}\cosh
\left[ \sqrt{z^{2}/a^{2}-m^{2}}(t^{\prime }-t)\right] \right\} ,
\]
where we have introduced notation
\begin{equation}
G_{\nu }(z,y)=I_{\nu }(y)\bar{K}_{\nu }(z)-\bar{I}_{\nu }(z)K_{\nu
}(y), \label{Geab}
\end{equation}
with the modified Bessel functions.

To obtain the Wightman function outside a single sphere on
background of the global monopole geometry, let us consider the
limit $b\rightarrow \infty $. In this limit the second integral on
the right of Eq. (\ref{unregWightab}) tends to zero (for large
values of the ratio $b/a$ the subintegrand is proportional to $
e^{-2bz/a}$), whereas the first integral and the term $r_{\nu
_l}[h(z)]$ do not depend on $b$. It follows from here that the
quantity
\begin{eqnarray}
\langle 0|\varphi (x)\varphi (x^{\prime })|0\rangle  &=&\frac{1}{%
2naS_{D}\sigma ^{D-1}}\sum_{l=0}^{\infty }\frac{2l+n}{(rr^{\prime })^{n/2}}%
C_{l}^{n/2}(\cos \theta )  \label{unregWightout} \\
&\times &\left\{ \int_{0}^{\infty
}\frac{zdz}{\sqrt{z^{2}+m^{2}a^{2}}}\frac{g_{\nu _{l}
}(z,zr/a)g_{\nu _{l} }(z,zr^{\prime }/a)}{\bar{J}_{\nu _{l}
}^{2}(z)+\bar{Y}_{\nu _{l}
}^{2}(z)}e^{i\sqrt{z^{2}/a^{2}+m^{2}}(t^{\prime }-t)}-r_{\nu _l
}[h(z)] \right\} \nonumber
\end{eqnarray}
is the Wightman function for the exterior region of a single
sphere with radius $a$. To find the boundary induced VEVs we have
to subtract the corresponding part for the unbounded monopole
geometry which, as we saw in previous section, can be presented in
the form (\ref{Mink}). Using the relation
\begin{equation}
\frac{g_{\nu }(z,zr/a)g_{\nu }(z,zr^{\prime }/a)}{\bar{J}_{\nu }^{2}(z)+\bar{%
Y}_{\nu }^{2}(z)}-J_{\nu }(zr/a)J_{\nu }(zr^{\prime }/a)=-\frac{1}{2}%
\sum_{s=1}^{2}\frac{\bar{J}_{\nu }(z)}{\bar{H}_{\nu }^{(s)}(z)}%
H_{\nu }^{(s)}(zr/a)H_{\nu }^{(s)}(zr^{\prime }/a) \label{relab}
\end{equation}
one obtains
\begin{eqnarray}
&&\langle 0|\varphi (x)\varphi (x^{\prime })|0\rangle -\langle 0_{{\rm m}%
}|\varphi (x)\varphi (x^{\prime })|0_{{\rm m}}\rangle =-\frac{1}{%
4nS_{D}\sigma ^{D-1}}\sum_{l=0}^{\infty }\frac{2l+n}{(rr^{\prime })^{n/2}}%
C_{l}^{n/2}(\cos \theta ) \nonumber \\
&&\times \left\{ \sum_{s=1}^{2}\int_{0}^{\infty }dz\,z\frac{e^{i\sqrt{%
z^{2}+m^{2}}(t^{\prime }-t)}}{\sqrt{z^{2}+m^{2}}}\frac{\bar{J}_{\nu _{l} }(za)}{%
\bar{H}_{\nu _{l} }^{(s)}(za)}H_{\nu _{l} }^{(s)}(zr)H_{\nu _{l}
}^{(s )}(zr^{\prime })+ \frac{2}{a}r_{\nu _l }[h(z)]\right\}.
\label{extdif}
\end{eqnarray}
In this formula we can rotate the integration contour on the right
by the angle $\pi /2$ for $s=1$ and by the angle $-\pi /2$ for
$s=2$. It can be easily seen that the residue terms coming from
the poles at the zeros of the functions $\bar{H}^{(s)}_{\nu }(za)$
cancel the second summand in the figure braces in Eq.
(\ref{extdif}). Further, the integrals over the segments $(0,ima)$
and $(0,-ima)$ of the imaginary axis cancel out and after
introducing the Bessel modified functions one obtains
\begin{eqnarray}
\langle \varphi (x)\varphi (x^{\prime })\rangle _{b}
&=&-\frac{1}{\pi n a
S_{D}\sigma ^{D-1}}\sum_{l=0}^{\infty }\frac{2l+n}{(rr^{\prime })^{n/2}}%
C_{l}^{n/2}(\cos \theta )  \label{regWightout} \\
&\times &\int_{ma}^{\infty }dz\, z\frac{\bar{I}_{\nu _{l}
}(z)}{\bar{K}_{\nu _{l} }(z)}\frac{K_{\nu _{l} }(zr/a)K_{\nu _{l}
}(zr^{\prime }/a)}{\sqrt{z^{2}-m^{2}a^2}}\cosh \! \left[
\sqrt{z^{2}/a^2-m^{2}}(t^{\prime }-t)\right] .  \nonumber
\end{eqnarray}
This formula differs from the one for the interior region by the
replacements $I_{\nu _{l} }\to K_{\nu _{l} }$, $K_{\nu _{l} }\to
I_{\nu _{l} }$. For the boundary induced VEV of the field square
in the outside region this leads to the formula
\begin{equation}
\langle \varphi ^{2}(x)\rangle _{b}=-\frac{1}{\pi ar^{n}S_{D}\sigma ^{D-1}}%
\sum_{l=0}^{\infty }D_{l}\int_{ma}^{\infty }dz\ z\frac{\bar{I}_{\nu _{l} }(z)}{%
\bar{K}_{\nu _{l} }(z)}\frac{K_{\nu _{l}
}^{2}(zr/a)}{\sqrt{z^{2}-m^{2}a^{2}}}, \label{regsquareout}
\end{equation}
where $D_{l}$ is defined in accord with Eq. (\ref{Dlang}). This
expression is finite for all values $r>a$ and diverges at the
sphere surface. The leading term in the corresponding asymptotic
expansion over $r-a$ has the form (\ref{epsinasimp}) with
replacement $a-r\to r-a$. In the case of the Dirichlet boundary
condition $\langle \varphi ^{2}(x)\rangle _{b}$ is monotonic
increasing negative function on $r$.

For the case of a massless scalar the asymptotic behavior of
boundary part (\ref{regsquareout}) at large distances from the
sphere can be obtained by introducing a new integration variable
$y=zr/a$ and expanding the subintegrand in terms of $a/r$. The
leading contribution for the summand with a given $l$ has an order
$(a/r)^{2\nu _{l} +D-1}$ (assuming that $A\neq \nu _{l} B$) and
the main contribution comes from the $l=0$ term. Using the value
for the standard integral involving the product of the functions
$K_{\nu }$ \cite{Prudnikov}, the leading term for the asymptotic
expansion over $a/r$ can be presented in the form
\begin{equation}\label{phi2rgga}
\langle \varphi ^{2}(x)\rangle _{b}\approx -\frac{\nu _0\Gamma
(2\nu _0+1/2)\Gamma (\nu _0+1/2)}{2^{2\nu _0+1}(a\sigma
)^{D-1}S_D\Gamma ^3(\nu _0+1)}\frac{A+B\nu _0}{A-B\nu _0}\left(
\frac{a}{r}\right) ^{2\nu _0+D-1}.
\end{equation}
Now comparing this with Eq. (\ref{renEMTpuremon}), we see that for
$\nu _0>0$ the VEV of the field square at large distances from the
sphere is dominated by the purely gravitational part. In the limit
$\sigma \ll 1$ the leading terms in the asymptotic expansion of
Eq. (\ref{regsquareout}) are derived by the way similar to that
used above for the inside region. For $\xi
>0$ the corresponding formula is obtained from Eq.
(\ref{phi2sigll1}) with replacements $a^2-r^2\to r^2-a^2$ under
the square root and $a/r \to r/a$ under the $\ln $ function. In
the case $\xi =0$ in Eq. (\ref{phi2sigll1b}) we have to make
replacements $I_{\nu _{l} }\to K_{\nu _{l} }$, $K_{\nu _{l} }\to
I_{\nu _{l} }$. In the right panel of Fig. \ref{fig1phi2} we have
presented the dependence of the boundary induced VEV
(\ref{regsquareout}) on $r/a$ for the region outside the sphere in
the case of a conformally coupled ($\xi =\xi _D $) massless
Dirichlet scalar in $D=3$ spatial dimensions. The curves
correspond to the values $\sigma =1$ (a), $\sigma =0.5$ (b),
$\sigma =0.2$ (c).

As in the interior case, the vacuum energy-momentum tensor is
diagonal and the corresponding components can be presented in the
form of a sum of purely gravitational and boundary parts, Eq.
(\ref{compdecomp}). The parts of these components induced by the
presence of a spherical shell are given by formulae
\begin{equation}
q_{b}(a,r)=-\frac{1}{2\pi a^{3}r^{n}S_{D}\sigma
^{D-1}}\sum_{l=0}^{\infty
}D_{l}\int_{ma}^{\infty }dz\,z^{3}\frac{\bar{I}_{\nu _{l} }(z)}{\bar{K}_{\nu _{l} }(z)}%
\frac{F_{\nu _{l} }^{(q)}\left[ K_{\nu _{l} }(zr/a)\right] }{\sqrt{z^{2}-m^{2}a^{2}}}%
,\quad q=\varepsilon ,\,p,\,p_{\perp },\quad r>a,  \label{q1out}
\end{equation}
where the functions $F_{\nu _{l} }^{(q)}\left[ f(y)\right] $ are
given by relations (\ref{Fineps}), (\ref{Finperad}) and
(\ref{Finpeaz}). As for the interior components, the quantities
(\ref{q1out}) diverge at the sphere surface, $r=a$. The leading
terms of the asymptotic expansions are
determined by same formulae (\ref{epsinasimp}) with the replacement $%
a-r\rightarrow r-a$. When $D=1$ from Eq. (\ref{q1out}) we obtain
the expectation values for the one dimensional semi-infinite
region $r>a$, given in Refs. \cite{Saha01,Rome02}.

For large distances from the sphere, $r\gg a$, the main
contribution into the VEV of the energy-momentum tensor comes from
the $l=0$ summand. The leading terms of the asymptotic expansions
have the form
\begin{equation}
q_{b}\approx -\frac{2^{-2\nu _{0}}a^{-D-1}\sigma ^{1-D}}{\pi \nu
_{0}S_{D}\Gamma ^{2}(\nu _{0})}\frac{A+B\nu _{0}}{A-B\nu
_{0}}\left( \frac{a }{r}\right) ^{2\nu _{0}+D+1}\int_{0}^{\infty
}dzz^{2\nu _{0}+2}F_{\nu _{0}}^{(q)}[K_{\nu _{0}}(z)],
\label{relfarDq}
\end{equation}
where $\nu _{0}$ is determined by formula (\ref{nu0}). Note that
the integrals in this formula can be evaluated using the value for
the integrals involving the product of the functions $K_{\nu }$
given in Ref. \cite{Prudnikov}. As we see, for $\nu _0>0$ and for
large distances from the sphere the vacuum energy-momentum tensor
is dominated by the purely gravitational part. In Fig.
\ref{fig2EMT} (right panel) we have presented the sphere induced
quantities $\varepsilon _b$, $p_b$, $p_{\perp b}$ for the region
outside a spherical shell as functions on $r/a$ in the case of a
conformally coupled massless $D=3$ Dirichlet scalar field on
background of the global monopole with the solid angle deficit
$\sigma =0.5$.

\section{Conclusion} \label{msec3:conc}

In the present paper we have investigated the Casimir densities
induced by a spherical shell on background of the
$D+1$-dimensional global monopole spacetime described by the
metric (\ref{mmetric}). The case of a massive scalar field with
general curvature coupling parameter and satisfying the Robin
boundary condition on the sphere is considered. All calculations
are made at zero temperature and we assume that the boundary
conditions are frequency independent. The latter means no
dispersive effects are taken into account. To obtain the
expectation values for the energy-momentum tensor we first
construct the positive frequency Wightmann function (note that the
Wightmann function is also important in considerations of the
response of a particle detector at a given state of motion
\cite{Birr82}). The application of the generalized Abel-Plana
formula to the mode sum over zeros of the corresponding
combinations of the cylinder functions allows us to extract the
part due to the global monopole geometry without boundaries and to
present the sphere induced part in terms of integrals which are
exponentially convergent in the coincidence limit at any strictly
interior or exterior points. The polarization of the scalar vacuum
by the global monopole without boundaries is investigated in a
number of previous papers \cite{Hisc90}--\cite{Beze01temp}, and
our main interest in this paper are the sphere induced parts for
the VEVs (Casimir densities). The expectation values for the
energy-momentum tensor are obtained by applying on the
corresponding Wightman function a certain second-order
differential operator and taking the coincidence limit. These
quantities diverge as the boundary is approached. Surface
divergences are well known in quantum field theory with boundaries
and are investigated near an arbitrary shaped smooth boundary.
They lead to divergent global quantities, such as the total energy
or vacuum forces acting on the sphere and additional
renormalization procedure is needed. The total Casimir energy for
a spherical shell on the global monopole background is
investigated in Refs. \cite{Bord96m,Beze99} using the zeta
function regularization method.

The boundary induced expectation values of the field square and
energy-momentum tensor for the region inside a spherical shell are
given by formulae (\ref{regsquare}) and (\ref{q1in}) respectively.
These expressions are finite at interior points and diverge on the
sphere surface. The leading term in the corresponding asymptotic
expansions is the same as for a sphere in the Minkowski bulk and
is zero for a conformally coupled scalar. The coefficients for the
subleading asymptotic terms will depend on the boundary curvature,
Robin coefficient, mass and the parameter $\sigma $ describing the
solid angle deficit. In section \ref{msec2:out} to deal with
discrete modes, first we consider the scalar vacuum in the
spherical layer between two surfaces. The quantities
characterizing the vacuum outside a single sphere are obtained
from this case in the limit when the radius of the outer sphere
tends to infinity. Subtracting the parts corresponding to the
space without boundaries, for the boundary induced parts of the
Wightman function and vacuum densities we derive formulae
(\ref{regWightout}), (\ref{regsquareout}) and (\ref{q1out}). They
differ from the ones for the interior region by the replacements
$I_{\nu _{l} }\rightarrow K_{\nu _{l} }$, $K_{\nu _{l}
}\rightarrow I_{\nu _{l} }$. Near the sphere centre and at large
distances from the sphere the VEVs for the field square and
energy-momentum tensor are dominated by the purely gravitational
parts corresponding to the monopole geometry without boundaries.
For the points near the sphere surface the boundary induced parts
in the VEVs dominate. We have also investigated the VEVs in the
limit of small values for the parameter $\sigma $, describing the
solid angle deficit, $\sigma \ll 1$. In this limit, corresponding
to strong gravitational fields, for $\xi >0$ the boundary induced
parts behave as  $\langle \varphi ^2 \rangle _b\sim \sigma
^{3/2-D}\exp (-\gamma /\sigma )$ and $\varepsilon _b\sim p_{\perp
b }\sim p_b/\sigma \sim \sigma ^{-D-1/2}\exp (-\gamma /\sigma )$,
with $\gamma =2\sqrt{n(n+1)\xi }|\ln (a/r)|$, and the boundary
induced stresses are strongly anisotropic. For a minimally coupled
scalar ($\xi =0$) and $\sigma \ll 1$ we have shown that $\langle
\varphi ^2 \rangle _b\sim q_b\sim \sigma ^{1-D}$, $q=\varepsilon
,p_{\perp },p $. As an illustration of general results, in Fig.
\ref{fig1phi2} and Fig. \ref{fig2EMT} we have plotted the sphere
induced VEVs of the field square and energy-momentum tensor
components as functions on $r/a$ for a conformally coupled
Dirichlet scalar in $D=3$ spatial dimensions. Note that in this
case the leading terms in the asymptotic expansions of the vacuum
energy-momentum tensor components near the boundary vanish. The
VEV of the field square is negative everywhere and the energy
density is negative/positive inside/outside the sphere.

\section*{Acknowledgement }

We acknowledge support from the Research Project of the Kurdistan
University. The work of AAS was supported in part by the Armenian
Ministry of Education and Science (Grant No.~0887).

\appendix

\section{On the zeros of the function $C^{ab}_{\nu }(\eta ,z)$}
\label{sec:app1}

Here we will first show that for real values of the coefficients
$A$, $B$, $A_b$, $B_b$ in (\ref{eigenmodes}) and (\ref{Fb}) all
roots of the equation
\begin{equation}
C_{\nu }^{ab}(\eta ,z)\equiv \bar{J}_{\nu }(z) \bar{Y}^{(b)}_{\nu
}(\eta z)-\bar{Y}_{\nu }(z)\bar{J}^{(b)}_{\nu }(\eta z )=0, \quad
\eta >1 \label{Cnuabap}
\end{equation}
are real or purely imaginary. To see this note that from the
relation $(C_{\nu }^{ab}(\eta ,z))^{*}=C_{\nu }^{ab}(\eta ,z^{*})$
it follows that if $z$ is a root to (\ref{Cnuabap}) when $z^{*}$
is also a root. Further, from the Bessel equation for the function
$g_{\nu } (z,z x)$ defined in (\ref{genu}) (considered as a
function on $x$) the following orthogonality relation can be
derived
\begin{equation}\label{orthcond}
  \int _{1}^{\eta }dx\, x g_{\nu }(z_1,z_1 x)g_{\nu }(z_2,z_2
  x)=0,\quad z_1^2\neq z_2^2,
\end{equation}
for two roots $z_1$ and $z_2$ of Eq. (\ref{Cnuabap}). Assuming
$z_1^{*2}\neq z_1^2$ and taking in this relation $z_2=z_1^{*}$, we
obtain a contradiction:
\begin{equation}\label{contr}
\int _{1}^{\eta }dx\, x |g_{\nu }(z_1,z_1 x)|^2=0,
\end{equation}
and, hence, $z_1^{*2}= z_1^2$. This means that the zeros of the
function $C_{\nu }^{ab}(\eta ,z)$ are real or purely imaginary.

In this paper we are interested in the limit $\eta \to \infty $.
Let us show that for large values $\eta $ all zeros are real. For
the purely imaginary $z$, $z=ye^{\pi i/2}$ with real $y$, the
function $C_{\nu }^{ab}(\eta ,z)$ can be expressed via the Bessel
modified functions:
\begin{equation}\label{Cnuimag}
C_{\nu }^{ab}(\eta ,ye^{\pi i/2})=\frac{2}{\pi }\left[
\bar{K}_{\nu }(y) \bar{I}^{(b)}_{\nu }(\eta y)-\bar{I}_{\nu
}(y)\bar{K}^{(b)}_{\nu }(\eta y )\right] .
\end{equation}
For large $\eta $ the contribution of the second term in the
square brackets is exponentially suppressed and one has
\begin{equation}\label{Cnuetall1}
C_{\nu }^{ab}(\eta ,ye^{\pi i/2})\approx \frac{2}{\pi }
\bar{K}_{\nu }(y) \bar{I}^{(b)}_{\nu }(\eta y),\quad \eta \gg 1.
\end{equation}
From the asymptotic expansion of the function $I_{\nu }(z)$ for
large values of the argument it can be seen that for large $\eta $
the function $\bar{I}^{(b)}_{\nu }(\eta y)$ has no zeros. Further,
using the recurrent relations for the Bessel modified functions
and expressions (\ref{eigenmodes}) for the coefficients $A$ and
$B$ with $n^1=1$, we can see that
\begin{equation}\label{Knubar}
\bar{K}_{\nu }(y)=-\frac{B_1}{a}\left[ \left( \nu
+\frac{n}{2}-\frac{A_1a}{B_1}\right) K_{\nu }(y)+yK_{\nu
-1}(y)\right] .
\end{equation}
Under the condition (\ref{nu0}) with $\nu \geq \nu _0$ the
expression in the square brackets of this equation is always
positive and the function $\bar{K}_{\nu }(y)$ has no real zeros.
As a result we conclude that for large $\eta $ the function
$C_{\nu }^{ab}(\eta ,ye^{\pi i/2})$ has no real zeros with respect
to $y$. Therefore, for sufficiently large $\eta $ all zeros of the
function $C_{\nu }^{ab}(\eta ,z)$ are real.

\end{document}